\documentstyle{article}
\input epsf
\begin{document}
\begin{titlepage}
April 1996	\hfill	CU-TP-747 
\vskip 1.2 in
\begin{center}
{\Large\bf Improvement of the Staggered Fermion Operators}
\\\vspace{2cm}
{\bf Yubing Luo} \\\vspace{1cm}
	Columbia University, Department of Physics, \\
	New york, NY 10027 \\
E - mail: roy@cuphyc.phys.columbia.edu \\
\end{center}
\vspace{2cm}

\begin{abstract} 
We present a complete and detailed derivation of the finite lattice
spacing corrections to staggered fermion matrix elements. Expanding
upon arguments of Sharpe, we explicitly implement the Symanzik
improvement program demonstrating the absence of order $a$ terms in
the Symanzik improved action. We propose a general program to
improve fermion operators to remove $O(a)$ corrections from 
their matrix elements, and demonstrate this program for the examples 
of matrix elements of fermion bilinears and $B_K$. We find the former 
does have $O(a)$ corrections while the latter does not.
\end{abstract}
\end{titlepage}

\section{Introduction}
With the ocuurence of a new generation of teraflop parallel
supercomputers, we will be able to simulate lattice QCD with 
smaller and smaller statistical errors. It is now of increased 
importance to gain control of various kinds of systematic errors
which either affect the numerical results directly or affect the 
way in which physical quantities are extracted. One of the most 
important systematic errors comes from the finite lattice 
spacing $a$ which generates errors of the order of 
$a{\Lambda}_{QCD}$. For the present lattice computation, 
this corresponds to the corrections of the 
order of $20\% \sim 30\%$. Another important systematic error 
comes from the choice of lattice operators. There exist a variety of
lattice operators which approach the same continuum operator in the
limit $a \to 0$. However, many of these operators differ from the 
continuum limit at $O(a)$, and a systematic formalism is needed to 
improve the lattice operators so as to remove these $O(a)$ 
corrections.

For the case of Wilson fermions, the standard lattice action differs
from the continuum quark action by a term of $O(a)$. So, both the 
action and the operators need corrections at order $a$. Applying 
the improvement program of Symanzik\cite{symanzik} to Wilson fermions,
a procedure was proposed in ref.\cite{wilson_improve:1} 
\cite{martinelli:91a} to reduce the systematic errors due to 
the finiteness of the lattice spacing, from terms of $O(a)$ to 
ones of $O(g_0^2a)$, and it was numerically demonstrated in 
ref.\cite{martinelli:91b} that this procedure can reduce the finite
$a$ corrections from $30\%$ to around $5\%$.

The meaning of the statement that there is no term of order $a$ in
the staggered fermion action is not clear. Let us
use the free staggered fermion action as an example:
\begin{equation}
    S_F = \sum_{x,\mu}a^4 \bar{\chi}(x) \eta_\mu(x)
    \frac{1}{2a}[\chi(x+\mu) -
    \chi(x-\mu)] + m\sum_xa^4\bar{\chi}(x) \chi(x).
\end{equation}
Following Golterman and Smit \cite{golterman},
we denote the Fourier components of the fields $\chi$ and
$\bar{\chi}$ as $\tilde{\chi}$ and $\tilde{\bar{\chi}}$, and
decompose momentum space as
\begin{equation}
    k = p + \pi_A,
\end{equation}
where $k_\mu \in (0, 2\pi/a)$, $p_\mu \in (0, \pi/a)$, $(\pi_A)_\mu
= A_\mu\pi$, in which $A_\mu = 0, 1$. 
If we define the fermion fields as
\begin{eqnarray}
    \tilde{\psi}(p)=\frac{1}{8}\sum_{A,B}(-1)^{A \cdot B}\gamma_A
    \tilde{\chi}(p+\pi_B), \\
    \tilde{\bar{\psi}}(p)=\frac{1}{8}\sum_{A,B}(-1)^{A \cdot B}
    \gamma_A^\dagger \tilde{\bar{\chi}}(p+\pi_B),
\end{eqnarray}
where
\begin{equation}
    \gamma_A =
    \gamma_1^{A_1}\gamma_2^{A_2}\gamma_3^{A_3}\gamma_4^{A_4},
\end{equation}
we can write the action as:
\begin{equation}\label{gs_action}
    S_F=\frac{(2\pi)^4}{\Omega} \sum_p \tilde{\bar{\psi}}(p)
    (\sum_\mu\gamma_\mu\frac{i}{a}\sin p_\mu a + m) \tilde{\psi}(p).
\end{equation}
where $\Omega$ is the lattice volume.
It is clear that there is no order $a$ term in the action, hence
the free staggered action is accurate to $O(a^2)$. However, the
coordinate fields corresponding to $\tilde{\psi}$
($\tilde{\bar{\psi}}$) are non-local superpositions of the
$\chi$'s ($\bar{\chi}$'s) over all the lattice sites.

On the other hand, if we define the local hypercubic fermion fields 
as in ref.\cite{kluberg-stern}
\begin{eqnarray}
    q(y)=\frac{1}{8}\sum_A\gamma_A\chi(y+A) = \frac{1}{2}
     	\sum_A\gamma_A\chi_A(y), \\
    \bar{q}(y)=\frac{1}{8}\sum_A\bar{\chi}(y+A)\gamma_A^\dagger = 
    	\frac{1}{2}\sum_A\bar{\chi}_A(y)\gamma_A^\dagger, 
\end{eqnarray}
where
\begin{equation}\label{y}
   x = y + A,	\qquad y_\mu = 0,\pm 2, ...
\end{equation}
then the action in the momentum space can be written as:
\begin{equation}
    S_F=\frac{(2\pi)^4}{\Omega} \sum_p \tilde{\bar{q}}(p)
    \{\sum_\mu[(\gamma_\mu \otimes I) \frac{i}{2a}\sin p_\mu2a
    +a(\gamma_5 \otimes \xi_{5\mu})(\frac{1}{a}\sin p_\mu a)^2] + m
    \}\tilde{q}(p).
\end{equation}
It is obvious that there are order $a$ terms in the action and in
the propagator for the fields $\tilde{q}$ and $\tilde{\bar{q}}$. In
this case, we say that the fields $q$ and $\bar{q}$ need to be
improved. There exists a set of improved fields
\begin{eqnarray}
    \chi_A^I(y)=(1-a\sum_\mu A_\mu \partial_\mu^L)\chi_A(y), \\
    \bar{\chi}_A^I(y) = \bar{\chi}_A(y)(1-a\sum_\mu A_\mu
    	\stackrel{\scriptstyle{\leftarrow}\scriptstyle{L}} 
    	{\partial_\mu}),
\end{eqnarray}
where
\begin{equation}
    \partial_\mu^Lf(y)=\frac{1}{4a}[f(y+2\mu)-f(y-2\mu)],
\end{equation}
such that
\begin{equation}
    S_F=\frac{(2\pi)^4}{\Omega} \sum_p \tilde{\bar{q}}^I(p)
    (\sum_\mu\gamma_\mu\frac{i}{a}\sin p_\mu a + m) \tilde{q}^I(p)
    + O(a^2).
\end{equation}
Note those improved fields are still local and superior to the
nonlocal fields both computationally and theoretically when gauge
couplings are included.
So, if we use the improved fields which remove the order $a$ terms 
from the action to construct a
lattice fermion operator, there will be no $O(a)$ corrections to its
free field matrix elements. For Landau gauge,
Sharpe \cite{sharpe:b, sharpe:c} 
proposed the following smeared operator:
\begin{equation}
    \chi_A(y)^{smeared} = \frac{1}{4}
    \sum_{\nu}\chi_A(y+2\nu[1-2A_\nu]).
\end{equation}
It is easy to show that
\begin{equation}
    \chi_A(y)^{smeared} = \chi_A^I(y) + O(a^2).
\end{equation}

The full staggered fermion action including gauge couplings is 
much more complicated. In section 2, we will give a set of improved
fermion field variables in terms of which the action 
has no explicit order $a$ terms at tree level. In section 3,
we will expand upon the argument given by Sharpe 
in ref \cite{sharpe:94a}
to prove that there are no $O(a)$ terms which can be
added to the staggered fermion action. Based on these two arguments,
we conclude that staggered fermion action is already accurate to
$O(a^2)$, and that we should use the improved field variables to 
construct fermion operators to reduce order $a$ corrections from 
their matrix elements. We apply 
this program to the case of $<0|\bar{s}\gamma_{54}d|K^0>$ and 
$B_K$ as examples.
We will also determine the additional operators that must be added
to improve the standard staggered fermion currents to define
operators whose matrix elements are accurate to $O(a^2)$.
We list the lattice symmetry transformation properties of the
fermion fields needed in this paper in the Appendix A.

\section{Improving the fermion fields}
Working with the even position variable $y$ of Eq.(\ref{y}), 
the lattice gauge-covariant derivative is defined as follows:
\begin{equation}
    D_\mu^Lf(y)=\frac{1}{4a}[U_\mu(y)U_\mu(y+\mu)f(y+2\mu) -
    U_\mu^{\dagger}(y-\mu)U_\mu^{\dagger}(y-2\mu)f(y-2\mu)].
\end{equation}
Following ref.\cite{kluberg-stern}, 
we define the gauge covariant fermion fields as
\begin{eqnarray} \label{cov_field}
    \varphi_A(y)={\cal{U}}_A(y)\chi_A(y), \nonumber  \\
    {\bar{\varphi}}_A(y)={\bar{\chi}}_A(y){\cal{U}}^\dagger_A(y),
\end{eqnarray}
where ${\cal{U}}_A(y)$ is the average of link products along 
the shortest paths from $y$ to $y+A$.
The classical continuum limit of action can be written as:
\begin{eqnarray}
    \lefteqn{S_F[\varphi, \bar{\varphi}] \stackrel{a\rightarrow0}
    {\longrightarrow} \int_y \sum_{AB} \bar{\varphi}_A(y) 
    \{ \sum_{\mu} \overline{(\gamma_\mu \otimes I)}_{AB} D_\mu 
    + m\delta_{AB} -a [ \sum_\mu \overline{(\gamma_5 \otimes 
     \xi_{5\mu})}_{AB}D^2_\mu}
    	\nonumber \\
   & & +\frac{1}{4} \sum_{\mu\nu}(\overline{(\gamma_\mu-\gamma_\nu)
       \otimes I} + \frac{i}{2}g_0\overline{\gamma_5 [\gamma_\mu,
       \gamma_\nu] \otimes \xi_5(\xi_\mu + \xi_\nu)})_{AB}
       F_{\mu\nu}(y)]\} \varphi_B(y)	\nonumber \\
   & & +O(a^2).
\end{eqnarray}
where $D_\mu=\partial_\mu+igA_\mu$ is the continuum covariant
derivative. At first sight, the action contains order $a$ terms.
Likewise, it is clear that the fermion propagator for the 
hypercubic fields $\varphi$ and $\bar{\varphi}$ deviates from the
continuum propagator by terms of order $a$.
However, if we introduce the following improved field variables
\begin{eqnarray}
    \chi^I_A(y) = (1-a\sum_\nu A_\nu D^L_\nu)\chi_A(y),	\\
    \bar{\chi}^I_A(y)=\bar{\chi}_A(y)(1-a\sum_\nu A_\nu 
    	\stackrel{\scriptstyle{\leftarrow}
    \scriptstyle{L}} {D_\nu}),
\end{eqnarray}
and replace $\chi$, $\bar{\chi}$ in Eq.\ref{cov_field} by 
$\chi^I$ and $\bar{\chi}^I$,
then the classical continuum limit of the action can be written as
\begin{equation}
    S_F[\varphi^I, \bar{\varphi}^I] =
     \int_y \sum_{AB} \bar{\varphi}^I_A(y) 
     [ \sum_\mu \overline{(\gamma_\mu \otimes I)}_{AB} D_\mu +
     m\delta_{AB}] \varphi^I_B(y) + O(a^2),
\end{equation}
We see the Feynman rules of the improved fields differ from those of
the continuum theory by terms of $O(a^2)$.
In ref.\cite{martinelli:91b} 
it was proven that if the matrix elements of the tree
level improved operators differ from the continuum ones only by terms
which are of $O(a^2)$, then in the full theory there are no terms of 
$O(a)$ nor terms of $O(ag_0^{2n}{\ln}^na)$. 
We restate their argument here.
Since the lowest order correction to the propagator and vertices are
of order $a^2$, a term proportional to $ag_0^{2n}{\ln}^na$
from an n-loop diagram can only occur if one of the loop
integrals diverges like $1/a$. If this is true, there will be at
most $n-1$ logarithmically divergent loop integrals which will
result in a term of $O(ag_0^{2n}{\ln}^{n-1}a)$ which behaves like 
$O(g_0^2a)$ as $a\rightarrow 0$. So
we conclude that there are no terms of $O(ag_0^{2n}{\ln}^na)$,
and therefore the matrix elements differ from their continuum ones
by terms at most of $O(g_0^2a)$.

Using the new fermion fields, we can construct improved fermion
operators. For example, the improved fermion bilinears have the
following form:
\begin{eqnarray} \label{improved_bilinear}
    \lefteqn{\bar{\chi}^I_A(y) \overline{(\gamma_S 
     \otimes \xi_F)}_{AB} \chi^I_B(y) = 
     \bar{\chi}_A(y)\overline{(\gamma_S \otimes \xi_F)}_{AB} 
     \chi_B(y)}		\nonumber	\\
    & & -\frac{a}{2}\sum_\nu \partial^L_\nu [\bar{\chi}_A(y)
       (\overline{(\gamma_S \otimes \xi_F)}_{AB} 
        - \overline{(\gamma_{5\nu S} \otimes \xi_{5\nu F})}_{AB})
        \chi_B(y)]	\nonumber	\\
    & & -\frac{a}{2}\sum_\nu \bar{\chi}_A(y)
       [\overline{(\gamma_{5\nu S} \otimes \xi_{5\nu F})}_{AB}
        -\overline{(\gamma_{S5\nu} \otimes \xi_{F5\nu})}_{AB}]
       D^L_\nu \chi_B(y)	\nonumber	\\
   & & +O(a^2).
\end{eqnarray}

\section{Improving the staggered fermion action}
In contrast to the calculation of matrix elements, the action is
already accurate through order $a$ to all orders in $g_0^2$, 
as we will now discuss. Thus
physical quantities that depend only on the form of the action (for
example, particle masses determined from correlation functions) will
have no corrections of order $g_0^{2n}a$. This can be demonstrated
by recognizing that if there were a correction of order $g_0^2a$, we
must necessarily be able to add some dimension-5 operators 
$a\sum_i c_ig_0^2O^{(5)}_i$ which must be invariant under the
lattice symmetry transformations to cancel this order
$g_0^2a$ correction(see ref.\cite{symanzik} \cite{sharpe:94a}
\cite{luscher}). However, following ref.\cite{sharpe:94a}, 
we will now prove that there exists no dimension-5 operator
(for the definition of the dimension of lattice operators, see
ref. \cite{luscher}) which is invariant under
the lattice symmetry group (rotations, axis reversal, translations,
$U(1) \otimes U(1)$, charge conjugation), and therefore, 
no order $a$ term can be added to the staggered fermion
action.

Following standard notation, we rewrite the staggered fermion action
as
\begin{eqnarray}\label{sec2:action} 
S_f = (2a)^4 \sum_{y,y'}\sum_{A,B}\bar{\chi}_A(y)^a [\sum_{\mu}
	\overline{(\gamma_\mu \otimes I)}_{AB} D_\mu(y,y')_{BC} 
		\nonumber  \\
      + m\delta(y-y')\delta_{AC}]\chi_C(y')^b 
      		U(y+A, y'+C)^{ab},
\end{eqnarray}
where
\begin{equation}
 (\overline{\gamma_S \otimes \xi_F})_{AB}
    = \frac{1}{4}Tr(\gamma_A^\dagger \gamma_S \gamma_B
    \gamma_F^\dagger),
\end{equation}
\begin{equation}
  D_\mu(y,y')_{AB} = \overline{D}_\mu(y,y')\delta_{AB} +
	a\overline{\Delta}_\mu(y,y')\overline{(\gamma_{\mu5} \otimes 
	\xi_{\mu5})}_{AB}, 
\end{equation}
in which
\[ \overline{D}_\mu(y,y') = \frac{1}{4a}[\delta(y+2\mu-y') - 
	\delta(y-2\mu-y')], \]
\[ \overline{\Delta}_\mu(y,y') = \frac{1}{4a^2}[\delta(y+2\mu-y') +
	\delta(y-2\mu-y') -2\delta(y-y')], \]
For convenience, we will not write out the $SU(3)$ links explicitly 
in the remainder of this section unless there would otherwise 
be confusion.  Given an operator, the reader can write out the 
full form very easily.  For example, starting with the operator
$\sum_\mu\bar{\chi}\overline{ \gamma_5 \otimes
\xi_{5\mu}}D^2_\mu \chi $ we would construct 
the corresponding gauge invariant operator by the
following substitution:
\begin{eqnarray}
 \sum_{y',y''}\sum_\mu \bar{\chi}_A(y) \overline{(\gamma_5 \otimes
	\xi_{5\mu})}_{AB} D_\mu(y,y')_{BC}D_\mu(y'y'')_{CD} \chi_D(y'')
     \longrightarrow \nonumber \\
     \sum_{y',y''}\sum_\mu \bar{\chi}_A(y)
	\overline{(\gamma_5 \otimes \xi_{5\mu})}_{AB} U(y+A,y+B)
	D_\mu(y,y')_{BC} U(y'+B,y''+C) \nonumber \\
	\times D_\mu(y',y'')_{CD} U(y'+C,y''+D) \chi_D(y'').
\end{eqnarray}
where $U(y+A, y+B)$ is the average of 
the products of link matrices corresponding to each of
the shortest paths from
point $y+A$ to $y+B$.

Using the transformation properties of the staggered fermion action 
(see Appendex A in detail), we try to construct all symmetrical
dimension-5 operators which have the general form $\bar{\chi}
\overline{\gamma_S \otimes \xi_F} f(D,\overline{D}) \chi$ where
$f$ is a homogenous real polynimial of degree 2.

Invariance under $U_A(1)$ requires that $S+F$ is odd, so only the
following combinations of $S \otimes F$ are valid:
\begin{equation}
	(I, \gamma_5, \gamma_{\mu\nu}, \gamma_{5\mu\nu}) \otimes 
	(\xi_\lambda, \xi_{5\lambda}),
\end{equation}
\begin{equation}
	(\gamma_\mu, \gamma_{5\mu}) \otimes 
	(I, \xi_5, \xi_{\lambda\tau}, \xi_{5\lambda\tau}).
\end{equation}

Under reflection with respect to a hyperplane normal to the $\rho$
direction, we have the following transformation:
\begin{eqnarray}
 	\chi \rightarrow {\cal{I}}_\rho\chi, \\
 	\bar{\chi} \rightarrow \bar{\chi}{\cal{I}}^{-1}_\rho,
\end{eqnarray}
and
\begin{eqnarray}
    \overline{D}_\mu \rightarrow (1-2\delta_{\mu\rho}){\cal{I}}_\rho
		\overline{D}_\mu {\cal{I}}^{-1}_\rho, \\
	\overline{\Delta}_\mu \rightarrow {\cal{I}}_\rho 
		\overline{\Delta}_\mu {\cal{I}}^{-1}_\rho,	\\
    D_\mu \rightarrow (1-2\delta_{\mu\rho}){\cal{I}}_\rho D_\mu 
		{\cal{I}}_\rho^{-1}.
\end{eqnarray}

Using the transformation formulae of $\gamma_S \otimes \xi_F$ listed
in Eq.(~\ref{Apend:2}) of the Appendix, we deduce that
axis reversal invariance and 
$U_A(1)$ invariance allow only the following terms:
\begin{equation}\label{term:1}
   \gamma_5 \otimes \xi_\mu D^2_\mu,
\end{equation}
\begin{equation}\label{term:2}
   \gamma_5 \otimes \xi_{5\mu} D^2_\mu,
\end{equation}
\begin{equation}\label{term:3}
   \gamma_5[\gamma_\mu, \gamma_\nu] \otimes \xi_5(\xi_\mu +
    	\xi_\nu) [D_\mu, D_\nu],
\end{equation}
\begin{equation}\label{term:4}
   \gamma_5[\gamma_\mu, \gamma_\nu] \otimes \xi_5(\xi_\mu -
    	\xi_\nu) \{D_\mu, D_\nu\}.
\end{equation}
where $D_\mu, D_\nu$ can be replaced by $\overline{D}_\mu,
\overline{D}_\nu$ without affecting these operators up to order $a^2$.

Under a rotation around the center of a hypercube, we have the 
following:
\begin{eqnarray}
    \chi \rightarrow {\cal{R}}^{(\rho\sigma)}\chi,	\\
    \bar{\chi}\rightarrow\bar{\chi}{\cal{R}}^{(\rho\sigma)-1} \\
    \overline{D}_\mu \rightarrow {\cal{R}}^{(\rho\sigma)}R_{\mu\nu}
    	\overline{D}_\nu {\cal{R}}^{(\rho\sigma)-1}	\\
    \overline{\Delta}_\mu\rightarrow {\cal{R}}^{(\rho\sigma)}
    	\left| R_{\mu\nu}\right|
    	\overline{\Delta}_\nu {\cal{R}}^{(\rho\sigma)-1}	\\
    D_\mu\rightarrow {\cal{R}}^{(\rho\sigma)}R_{\mu\nu}
    	D_\nu {\cal{R}}^{(\rho\sigma)-1}	\\
\end{eqnarray}

Combining the transformation properties listed in
Eq.(~\ref{Apend:3}) of the Appendix,
the rotational invariance will further eliminate the term in
Eq.(\ref{term:1}) but allows the remaining three terms
Eq.(\ref{term:2} - \ref{term:4}).

Finally, let's discuss invariances under translation by one 
lattice unit:
\begin{eqnarray}
    \chi \rightarrow {\cal{S}}^{(\rho)}\chi,	\\
    \bar{\chi} \rightarrow \bar{\chi}{\cal{S}}^{(\rho)-1}, \\
    \overline{D}_\mu \rightarrow {\cal{S}}^{(\rho)}
    	\overline{D}_\mu {\cal{S}}^{(\rho)-1},	\\
    \overline{\Delta}_\mu \rightarrow {\cal{S}}^{(\rho)}
    	\overline{\Delta}_\mu {\cal{S}}^{(\rho)-1},	\\
    D_\mu \rightarrow {\cal{S}}^{(\rho)}
    	\Lambda_{\gamma_\mu \otimes I}^{(\rho)-1}
    	D_\mu {\cal{S}}^{(\rho)-1},
\end{eqnarray}
where
\begin{eqnarray}
    \Lambda_{S\otimes F}^{(\rho)}(y,y')_{AB} = \varepsilon(F) 
    \left\{ (-1)^{F_\rho}\delta_{AB}\delta(y-y')
    +a[(-1)^{F_\rho}-(-1)^{S_\rho}] \right. \nonumber \\
    \left. \times [a\delta_{AB}\overline{\Delta}_\rho(y,y') +
    \overline{(\gamma_{5\rho} \otimes \xi_{5\rho})}_{AB}
    \overline{D}_\rho(y,y')] \right\},
    	\\
    \Lambda_{S\otimes F}^{(\rho)-1}(y,y')_{AB} = \varepsilon(F) 
    \left\{ (-1)^{F_\rho}\delta_{AB}\delta(y-y')
    +a[(-1)^{F_\rho}-(-1)^{S_\rho}] \right.  \nonumber \\
    \left. \times [a\delta_{AB}\overline{\Delta}_\rho(y,y') -
    \overline{(\gamma_{5\rho} \otimes \xi_{5\rho})}_{AB}
    \overline{D}_\rho(y,y')] \right\},
\end{eqnarray}
and
\begin{eqnarray}
    {\cal{S}}^{(\rho)-1} \overline{\gamma_S \otimes \xi_F}
    {\cal{S}}^{(\rho)} = \overline{\gamma_S \otimes \xi_F}
    \Lambda_{S \otimes F}^{(\rho)}.
\end{eqnarray}
From these properties, we can see that none of the terms listed in
Eq.(\ref{term:2} - \ref{term:4}) are invariant under 
lattice translation! So, we conclude that there is no dimension-5 
fermion operator which is invariant under the lattice symmetry group,
and therefore no dimension-5 operator can be added to the
staggered fermion action.

\section{Applications:}
As we argued above, actual numerical simulation should use the
improved fermion field variables. 
However, in most situations, we can use the
improvement program proposed in this paper to remove the $O(a)$
corrections without increasing the computational work. Here, we
apply this program to the calculation of the matrix element 
$<0|\bar{s}\gamma_{54}d|K^0>$ which gives $f_K$ in the continuum,
and the calculation of $B_K$. We will show that the former differs
from its continuum counterpart by $O(m_ka)$, but $B_K$ has no
$O(a)$ corrections. We also apply the improvement program to the 
matrix elements of lattice currents.

\subsection{$<0|\bar{s}\gamma_{54}d|K^0>$:} \label{application:1}
The axial current used in the (Landau gauge) numerical simulation is:
\begin{equation}
    A_\mu(y) = \sum_{AB}\bar{\chi}_A \overline{(\gamma_{5\mu} \otimes
    \xi_5)}_{AB} \chi_B(y).
\end{equation}
From the continuum expression
\begin{equation}
    P(t)^{cont} = <0|A_4(t)^{cont}|K^0> = \sqrt{2}f_K m_k
    e^{-m_k|t|},
\end{equation}
we define, on the lattice,
\begin{equation}
    P(t) = <0|\sum_{\vec{x}}A_4(\vec{x},t)|K^0>,
\end{equation}
and put the wall source that creates the $K^0$ on the time 
slice at $t=0$. Then we will have
\begin{eqnarray}
    P(t) = \left\{ \begin{array}{l}
	\sqrt{2}f_K^+ m_k e^{-m_k|t|}, \qquad (t>0)	\\
	\sqrt{2}f_K^- m_k e^{-m_k|t|}, \qquad (t<0)
	\end{array} \right .
\end{eqnarray}
where
\begin{equation}
    f_K^{\pm} = f_K \pm O(m_ka).
\end{equation}

If we don't consider $O(g_0^2 a)$ terms, we can take only the term
\[ -\frac{a}{2}\sum_\nu \partial_\nu^L [\bar{\chi}_A 
	\overline{(\gamma_{54} \otimes
	\xi_5)}_{AB}\chi_B]	\]
in Eq.(~\ref{improved_bilinear}) because other terms
contribute zero ``flavor'' trace at the tree level. So, we have
\begin{eqnarray}
    P(t)^{Imp} & = & P(t) - \frac{a}{2}\partial_4^L P(t) \nonumber \\
    & = & \left\{ \begin{array}{l}
        \sqrt{2}f_K^{+,Imp}m_k e^{-m_k|t|}, \qquad (t>0)	\\
	\sqrt{2}f_K^{-,Imp}m_k e^{-m_k|t|}, \qquad (t<0)
	\end{array} \right .
\end{eqnarray}
and
\begin{eqnarray}
    f_K^{+,Imp} = (1+\frac{1}{2}m_ka)f_K^+ + O(a^2), \\
    f_K^{-,Imp} = (1-\frac{1}{2}m_ka)f_K^- + O(a^2), \\
    f_K^{\pm,Imp} = f_K \pm O(a^2).
\end{eqnarray}
So, we get
\begin{equation}
    f_K^{\pm} = (1 \mp \frac{1}{2}m_ka)f_K + O(a^2).
\end{equation}

\begin{figure}[htb]
    \epsfxsize=120mm
    \epsfbox{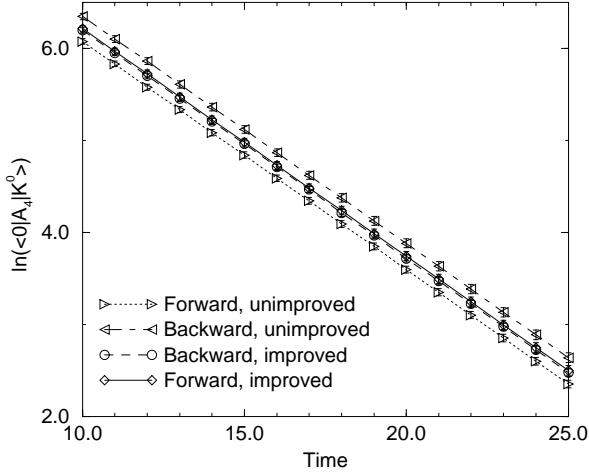}
    \caption{\label{fig.1} The value of ${\ln}(<0|A_4(t)|K^0>)$ with
    respect to the time $t$. Where $m_da = m_sa = 0.01$. Calculated 
    with the cubic wall source method.}
\end{figure}

The numerical data (from the full QCD simulation on a 
$16^3 \times 40$ lattice with the cubic wall source, at the quark
mass $m_sa = m_da = 0.01$, see ref.\cite{wjlee})
for the unimproved and improved matrix elements
are shown in Figure \ref{fig.1},
from which we see 
that $(f_K^- - f_K^+)/f_K
\approx m_ka \sim 25\%$ and $(f_K^{-,Imp} - f_K^{+,Imp})/f_K 
\sim 5\%$ is much smaller. 

From this simple example, we can see that if we do not consider
$O(g^2a)$ corrections, the improved operator is eqivalent to the
extrapolation:
\begin{equation}
    P(t) = \sqrt{2}f_K^{Imp} m_k e^{-m_k|t+1/2|},
\end{equation}

\subsection{$B_K$:}
The formula for calculating $B_K$ is:
\begin{equation}
    B_K = \frac{{\cal{M}}_K}{\frac{8}{3} {\cal{M}}_K^V},
\end{equation}
where
\begin{eqnarray}
    {\cal{M}}_K = <\overline{K^0}|\bar{s}\gamma_\mu (1+\gamma_5)d
    	\bar{s}\gamma_\mu (1+\gamma_5)d|K^0>,	\\
    {\cal{M}}_K^V= <\overline{K^0}|\bar{s}\gamma_4\gamma_5 d|0> 
    	<0|\bar{s}\gamma_4\gamma_5 d|K^0>.
\end{eqnarray}
The improved numerator is (omitting the $O(g_0^2a)$ terms):
\begin{equation}
    {\cal{M}}_K^{Imp} = {\cal{M}}_K - \frac{a}{2} \partial_4^L
        {\cal{M}}_K + O(a^2).
\end{equation}
Since ${\cal{M}}_K(t)$ is a plateau (i.e. time independent)
 within the statistical error,
there is no $O(a)$ corrections to the numerator.

Since
\begin{equation}
    f_K^+ f_K^- = f_K^2 + O(a^2),
\end{equation}
the denominator ${\cal{M}}_k^V$ also has no $O(a)$ corrections
even if no attention is paid to an accurate definition of $f_K$.
Hence, we showed that there is neither $O(a)$ 
nor $O(ag_0^{2n}\log^na)$ corrections to $B_K$.
Sharpe \cite{sharpe:94a} has examined this question in greater detail
and argued that in fact there are no corrections of $O(g_0^{2n}a)$
also.  However, if we calculated the denominator 
only in one time direction and took the square of
 $f_K^+$ (or $f_K^-$), there would be an error of
order of $O(m_ka)$.

\subsection{Renormalization of lattice currents:}
The lattice currents can be written as
\begin{equation}
    J^F_{latt} = \bar{\chi} \overline{\gamma_J \otimes
    		\xi_F} \chi,
\end{equation}
and according to ref\cite{currents},
their renormalized continuum forms can be written as
\begin{equation}
    J^F_{cont} = Z_J \kappa^F_J J^F_{latt},
\end{equation}
where $Z_J$ is the usual (divergent) renormaliation constant and
$\kappa^F_J$ is a finite lattice renormalization constant. 
Using the method developed in this paper, we can explicitly
determine the improved currents accurate to $O(a^2)$. For example,
the conserved vector current and axial vector current corresponding
to the $U_V(1) \otimes U_A(1)$ lattice symmetry can be written as
follows:
\begin{eqnarray}
    V_\mu^I(y) = V_\mu(y) -\frac{a}{2}\sum_\nu \partial_\nu 
    		[\bar{\chi}_A (\overline{\gamma_\mu \otimes
    		I})_{AB} \chi_B]  \nonumber \\ \qquad
    		-\frac{a}{2}\partial_\mu 
    		[\bar{\chi}_A (\overline{\gamma_\mu \otimes
    		I})_{AB} \chi_B]  \nonumber \\ \qquad
    		-\frac{a}{4}\sum_\nu \partial_\nu 
    		[\bar{\chi}_A (\overline{\gamma_{5[\mu,\nu]} 
    		\otimes \xi_{5\nu}})_{AB} \chi_B] + O(a^2),
\end{eqnarray}
\begin{eqnarray}
    A_\mu^{\xi_5,I}(y) = A_\mu^{\xi_5}(y) 
    		-\frac{a}{2}\sum_\nu \partial_\nu 
    		[\bar{\chi}_A (\overline{\gamma_{\mu 5} \otimes
    		\xi_5})_{AB} \chi_B]  \nonumber \\ \qquad
    		-\frac{a}{2}\partial_\mu 
    		[\bar{\chi}_A (\overline{\gamma_{\mu 5} \otimes
    		\xi_5})_{AB} \chi_B]  \nonumber \\ \qquad
    		+\frac{a}{4}\sum_\nu \partial_\nu 
    		[\bar{\chi}_A (\overline{\gamma_{[\mu,\nu]} 
    		\otimes \xi_{\nu}})_{AB} \chi_B] + O(a^2).
\end{eqnarray}
The effect of the second term on the right hand side is to shift the
position y, labeling the current, from the corner to the center of
the hypercube. The third term whose effect is to shift in the
$\mu$'s direction occurs here because the currents are non-local
operators which involve an overlap between two nearest
hypercubes. The forth term is a mixing of a different spin-flavor
operator and is necessary to remove all order $a$ effects from a
general matrix element.

\section{Summary}
In this paper, based on the demonstration of that there is no
dimension-5 fermion operator which is invariant under all lattice
symmetry transformations and that there exists a set of improved
fermion fields with respect to which 
the tree-level action has no order $a$ terms, 
we concluded that the staggered fermion
action is already in fact improved to $O(a^2)$.
We argued that to remove order $a$ corrections from the matrix
elements, the first step is to use the proposed improved
fermion field variables
to construct fermion operators so that they differ from the continuum 
ones by order of $a^2$ at the tree level. Then we showed that to all
orders of perturbation theory treating $g_0^2\ln{a} \sim O(1)$,
the correction is at most of
$O(g_0^2a)$. Furthermore, for the on-shell quantities, they are
accurate to $O(a^2)$.  We applied our program to the matrix 
element $<0|\bar{s}\gamma_{54}d|K^0>$ and found that 
the unimproved one differs from the continuum by a factor of $O(m_ka)$. 
At the same time, we showed that there is no $O(a)$ corrections to 
$B_K$, which is consistent with the result of Sharpe \cite{sharpe:94a}.
We also discussed the matrix elements of the lattice currents, and
obtained the explicit terms which should be added to the original
current operators to define improved operators accurate through $O(a)$.

\bigskip
\bigskip

\section{Acknowledgements}
I warmly thank Norman H. Christ who proposed this topic, stimulated
me to finish this paper, and gave me the benifit of
extensive discussions during every stage of this work.
I am also grateful to Weonjong Lee for the numerical data 
appearing in this paper and many helpful discussions. I
also thank Bob Mawhinney for the helpful and interesting discussions
on the lattice symmetry of staggered fermions.
Adrian Kahler and ChengZhong Sui checked out some typing mistakes,
their kind hospitality is gratefully acknowledged.

\appendix
\begin{appendix}
\section{Symmetry Properties of Staggered Fermion}
For completeness, we collect some formulae connected with the
transformation properties of the staggered fermion under the lattice
symmetry group from ref.\cite{morel}.
\subsection{$U(1)_A$:}
\begin{eqnarray}
	\chi_A(y) \rightarrow e^{i\alpha\varepsilon(A)} \chi_A(y) \\
	\bar{\chi}_A(y) \rightarrow e^{i\alpha\varepsilon(A)}
		\bar{\chi}_A(y)
\end{eqnarray}
where
\[	\varepsilon(A) = (-1)^{\sum_{\mu}A_\mu}	\]

\subsection{Reflection with respect to a Hyperplane:}
\[ I_H^\rho:\quad \left\{ \begin{array}{l}
	x'_\rho = -x_\rho + 1,	\\
	x'_\mu = x_\mu \qquad \qquad \mu \not= \rho
	\end{array}	\right.	\]

The transformation of the fermion fields are:
\begin{eqnarray}
	\chi_A(y) \rightarrow \sum_{B}\sum_{y'} {\cal{I}}_\rho
	(y,y')_{AB}\chi_B(y'),	\\
	\bar{\chi}_A(y) \rightarrow \sum_{B}\sum_{y'}\bar{\chi}_B(y')
	{\cal{I}}_\rho^{-1}(y',y)_{BA},
\end{eqnarray}
where
\begin{eqnarray}
	{\cal{I}}_\rho(y,y')_{AB} = \overline{(\gamma_{\rho5} \otimes
	\xi_5)}_{AB} \delta(I_{\rho}y-y'),	\\
	{\cal{I}}_\rho^{-1}(y,y')_{AB} = \overline{(\gamma_{5\rho} 
	\otimes \xi_5)}_{AB} \delta(I_{\rho}y-y'),
\end{eqnarray}
and
\[	({I_\rho}y)_\mu = \left\{\begin{array}{l}
		y_\mu, \qquad (\mu \not= \rho) \\
		-y_\rho, \qquad (\mu = \rho) 
		\end{array} \right.	\]
The spin-flavor matrices transform as:
\begin{equation}\label{Apend:2}
    \overline{(\gamma_S \otimes \xi_F)} \rightarrow
    \overline{(\gamma_{\rho5} \otimes \xi_5)}
    \overline{(\gamma_S \otimes \xi_F)}
    \overline{(\gamma_{5\rho} \otimes \xi_5)},
\end{equation}
which are explicitly listed as following:
\[ \begin{array}{ll}
    S & \qquad F  \\
    I \rightarrow I & \qquad I \rightarrow I  \\
    \gamma_5 \rightarrow -\gamma_5 & \qquad \xi_5 \rightarrow \xi_5  \\
    \gamma_\mu \rightarrow (1-2\delta_{\mu\rho})\gamma_\mu & \qquad
    	\xi_\mu \rightarrow -\xi_\mu  \\
    \gamma_{5\mu} \rightarrow -(1-2\delta_{\mu\rho})\gamma_{5\mu} & \qquad 
    	\xi_{5\mu} \rightarrow -\xi_{5\mu}  \\
    \gamma_{\mu\nu} \rightarrow (1-2\delta_{\mu\rho}) 
    	(1-2\delta_{\nu\rho})\gamma_{\mu\nu} & \qquad 
    	\xi_{\mu\nu} \rightarrow \xi_{\mu\nu}  \\
    \gamma_{5\mu\nu} \rightarrow -(1-2\delta_{\mu\rho}) 
    	(1-2\delta_{\nu\rho})\gamma_{5\mu\nu} & \qquad
    	\xi_{5\mu\nu} \rightarrow \xi_{5\mu\nu} 
\end{array}	\]

\subsection{Rotations by $\pi/2$ arround the center of a hyperplane:}
\[ R_H^{(\rho\sigma)}:\quad \left\{ \begin{array}{l}
	x'_\rho = x_\sigma, \\
	x'_\sigma = -x_\rho+1,	\\
	x'_\mu = x_\mu, \qquad (\mu \not= \rho, \sigma)	\\
	\end{array}	\right.	\]

Define $R_{\mu\nu}$ such that
\[ (Ry)_\mu = R_{\mu\nu}y_\nu = \left\{\begin{array}{l}
	y_\sigma, \qquad (\mu=\rho)	\\
	-y_\rho,\qquad (\mu=\sigma)	\\
	y_\mu, \qquad (\mu \not= \rho, \sigma)
	\end{array} \right.	\]

Then, we have the following transformation:
\begin{eqnarray}
    \chi_A(y) \rightarrow \sum_{B}\sum_{y'} {\cal{R}}^{(\rho\sigma)}
    	(y,y')_{AB} \chi_B(y'),	\\
    \bar{\chi}_A(y) \rightarrow \sum_{B}\sum_{y'} \bar{\chi}_B(y') 
    	{\cal{R}}^{(\rho\sigma)}(y,y')_{AB}^{-1},
\end{eqnarray}
where
\begin{eqnarray}
    {\cal{R}}^{(\rho\sigma)}(y,y')_{AB}=\frac{1}{2} 
    	\overline{((1-\gamma_{\rho\sigma}) \otimes
    	(\xi_\sigma-\xi_\rho))}_{AB}\delta(R^{-1}y-y'),	\\
    {\cal{R}}^{(\rho\sigma)}(y,y')_{AB}^{-1}=\frac{1}{2} 
    	\overline{((1+\gamma_{\rho\sigma}) \otimes
    	(\xi_\sigma-\xi_\rho))}_{AB}\delta(Ry-y'),
\end{eqnarray}
The spin-flavor matrices transform as:
\begin{equation}\label{Apend:3}
    \overline{(\gamma_S \otimes \xi_F)} \rightarrow
    \frac{1}{2}\overline{((1-\gamma_{\rho\sigma})\otimes
    (\xi_\sigma-\xi_\rho))} \overline{(\gamma_S \otimes \xi_F)}
    \frac{1}{2}\overline{((1+\gamma_{\rho\sigma}) \otimes
    	(\xi_\sigma-\xi_\rho))}.
\end{equation}
which are explicitly listed as following:
\[ \begin{array}{ll}
    S & \qquad F  \\
    I \rightarrow I & \qquad I \rightarrow I  \\
    \gamma_5 \rightarrow \gamma_5 & \qquad \xi_5 \rightarrow -\xi_5  \\
    \gamma_\mu \rightarrow R_{\mu\lambda}\gamma_\lambda & \qquad
    	\xi_\mu \rightarrow -|R_{\mu\lambda}|\xi_\lambda  \\
    \gamma_{5\mu} \rightarrow R_{\mu\lambda}\gamma_{5\lambda} & \qquad
    	\xi_{5\mu} \rightarrow |R_{\mu\lambda}|\xi_{5\lambda}  \\
    \gamma_{\mu\nu} \rightarrow R_{\mu\lambda}R_{\nu\tau}
    	\gamma_{\lambda\tau} & \qquad 
    	\xi_{\mu\nu} \rightarrow |R_{\mu\lambda}||R_{\nu\tau}|
    	\xi_{\lambda\tau}  \\
    \gamma_{5\mu\nu} \rightarrow R_{\mu\lambda}R_{\nu\tau}
    	\gamma_{5\lambda\tau} & \qquad 
    	\xi_{5\mu\nu} \rightarrow -|R_{\mu\lambda}||R_{\nu\tau}|
    	\xi_{5\lambda\tau}
\end{array}	\]

\subsection{Translations by one lattice unit:}
\[  T_\rho: \quad \left\{ \begin{array}{l}
	x'_\rho = x_\rho+1,	\\
	x'_\mu = x_\mu, \qquad (\mu \not= \rho)
	\end{array} \right.	\]
lead to the following transformation on the $\chi_A$,
$\bar{\chi}_A$'s:
\begin{eqnarray}
    \chi_A(y) \rightarrow \sum_{B}\sum_{y'} {\cal{S}}^{(\rho)}
    	(y,y')_{AB} \chi_B(y'),	\\
    \bar{\chi}_A(y) \rightarrow \sum_{B}\sum_{y'} \bar{\chi}_B(y') 
    	{\cal{S}}^{(\rho)}(y,y')_{AB}^{-1},
\end{eqnarray}
where
\begin{eqnarray}
    {\cal{S}}^{(\rho)}(y,y')_{AB}=\frac{1}{2} 
    	[(\overline{I\otimes\xi_\rho}-
    	\overline{\gamma_{\rho5}\otimes\xi_5})_{AB}\delta(y-y')
    	\nonumber\\ +(\overline{I\otimes\xi_\rho}+
    	\overline{\gamma_{\rho5}\otimes\xi_5})_{AB}\delta(y+2\rho-y')],
			\\
    {\cal{S}}^{(\rho)}(y,y')_{AB}^{-1}=\frac{1}{2} 
    	[(\overline{I\otimes\xi_\rho}-
    	\overline{\gamma_{\rho5}\otimes\xi_5})_{AB}\delta(y-2\rho-y')
    	\nonumber \\ +(\overline{I\otimes\xi_\rho}+
    	\overline{\gamma_{\rho5}\otimes\xi_5})_{AB}\delta(y-y')],
\end{eqnarray}

\end{appendix}

\end{document}